\documentclass[letter,12pt]{article}
\usepackage[T1]{fontenc}
\usepackage[dvips]{graphicx}
\graphicspath{{images/}}
\usepackage{verbatim}
\usepackage{amsmath,amsthm}
\usepackage{xspace}
\usepackage{pifont}
\usepackage{graphicx}
\usepackage{amssymb}
\usepackage{epic, eepic}
\usepackage{dsfont}
\usepackage{amssymb}
\usepackage{makeidx}
\usepackage{mathrsfs}
\usepackage{exscale}
\usepackage{color} 
\usepackage{overpic} 
\usepackage{bm}
\usepackage{bbm}
\usepackage{booktabs} 
\usepackage{color, colortbl}
\usepackage{subcaption}

\definecolor{Gray}{gray}{0.9}

\usepackage{amsmath,afterpage}
\usepackage{epsf}
\usepackage{graphics,color}

\def\0{\mathbf{0}}

\def\lam{\lambda}
\def\rr{\rightarrow}

\def \< {\langle}
\def \> {\rangle}

\def\beqa{\begin{eqnarray}}
\def\eeqa{\end{eqnarray}}
\def\beqas{\begin{eqnarray*}}
\def\eeqas{\end{eqnarray*}}

\newtheorem{theorem}{Theorem}[section]

\newtheorem{proposition}[theorem]{Proposition}

\newtheorem{corollary}[theorem]{Corollary}

\newtheorem{remark}[theorem]{Remark}

\numberwithin{equation}{section}
\newcommand{\hatd}[1]{{}}

\newcommand{\bd}{\begin{displaymath}}
\newcommand{\ed}{\end{displaymath}}
\newcommand{\be}{\begin{equation}}
\newcommand{\ee}{\end{equation}}
\newcommand{\bq}{\begin{eqnarray}}
\newcommand{\eq}{\end{eqnarray}}
\newcommand{\bn}{\begin{eqnarray*}}
\newcommand{\en}{\end{eqnarray*}}

\newcommand{\re}{\mathds{R}}

\def\wt{\widetilde}

\def\one{\mathds{1}}

\usepackage{authblk}

\title{Optimal trading: the importance of being adaptive}
\author[1]{Claudio Bellani }
\author[1,2]{Damiano Brigo }
\author[1]{Alex Done}
\author[1,2]{Eyal Neuman \thanks{http://eyaln13.wixsite.com/eyal-neuman}}
\affil[1]{Department of Mathematics, Imperial College London }
\affil[2]{CFM-Imperial College Institute, London}

\begin{document}

 \vspace{-0.5cm}
\maketitle

\begin{abstract}
We compare optimal static and dynamic solutions in trade execution. An optimal trade execution problem is considered  where a trader is looking at a short-term price predictive signal while trading. When the trader creates an instantaneous market impact, 
it is shown that transaction costs of optimal adaptive strategies are substantially lower than the corresponding costs of the optimal static strategy. In the same spirit, in the case of transient impact it is shown that strategies that observe the signal a finite number of times can dramatically reduce the transaction costs and improve the performance of the optimal static strategy. 

\end{abstract}

\section{Introduction}
In this paper 	we answer a fundamental question in optimal execution: 

\bigskip

can we find relevant models showing a large improvement in expected trading cost plus risk when moving from optimal static solutions to optimal dynamic ones?

\bigskip
 
This problem is relatively original as there is almost no literature comparing the two classes of solutions in the same model. The problem is further complicated by the fact that, in the literature, at times the dynamic problem has been studied, whereas other times the static problem has been considered.  Furthermore, there are cases where even though the solution is sought in the dynamic class, it turns out to be static. For example, Bertsimas and Lo \cite{BLA98} seek the solution in the dynamic class, but this turns out to be static, unless an information signal is added to the price process. Almgren and Chriss \cite{OPTEXECAC00} seek the solution directly in the static class, due to tractability. Gatheral and Schied \cite{GatheralSchied} seek the solution in the dynamic class, and indeed it turns out to be non-static. 

The trading costs in execution problems stem from market impact. Market impact refers to the empirical fact that the execution of a large order affects the price of the underlying asset. Usually, this effect causes an unfavorable additional execution cost for the trader who is performing the exchange. As a result, a trader who wishes to minimize his trading costs has to split his order into a sequence of smaller orders which are executed over a finite time horizon. 
Academic efforts to reduce the transaction costs of large trades started with the seminal papers of Almgren and Chriss \cite{OPTEXECAC00} and Bertsimas and Lo \cite{BLA98}. Both models deal with 
the trading process of one large market participant (for instance an asset manager or a bank) who would like to buy or sell a large amount of shares or contracts during a specified duration. 
The cost minimization problem takes into account market impact
(see \cite{citeulike:13497373} and references therein) and therefore demands to trade slowly, or at least at a pace which takes into account the available liquidity. It is worth noticing that there are several types of market impact, including instantaneous, transient and permanent impact, and in this paper we will only consider instantaneous and transient impact.  
On the other hand, traders have an incentive to trade rapidly, because they do not want to carry the risk of an adverse price move far away from their decision price. The tradeoff between market impact and market risk is usually translated into a stochastic control problem where the trader's strategy (i.e. the control) is the trading speed or the amount inventory liquidated at any time within the time horizon. Loosely speaking, the optimal strategy minimizes the risk-cost functional over a certain class of strategies.   

More recent literature on optimal execution inlcudes Tucci and Vega \cite{tucci} who analyze optimal execution under linear and nonlinear impact, characterizing the related optimization as a quadratic problem. Gatheral et al. \cite{gatheralschiedslynko}  consider earlier works by Gatheral himself and Alfonsi and Schied on models combining nonlinear price impact with exponential decay of market impact, explaining why in some cases this leads to price manipulation while in other cases there is no such effect. Finally, Labadie and Lehalle \cite{labadie} derive explicit recursive formulas for target close and implementation shortfall in the Almgren-Chriss framework. They show how to add a minimum participation rate constraint and  study an alternative set of risk measures for the optimisation of algorithmic trading curves. This is done under a self-similar process and a new risk measure, the p-variation, is introduced and analyzed. 

As we hinted above, within the framework of optimal execution we usually distinguish between two classes of trading strategies: static (or deterministic) and adaptive (or dynamic). When seen from the initial time of the trade execution, static strategies are deterministic strategies that are completely decided at that time, based only on the information that is revealed to the trader at that initial time.   Adaptive strategies are instead random when seen from the initial time, in that they will  depend at each time point on the whole information that is available at that time. This models the fact that a trader will be able to react to new available information and adjust her strategy. Technically, adaptive strategies are stochastic processes that are adapted to the relevant market information filtration in the given  model.  Clearly the class of static strategies is a subset of the class of adaptive strategies, therefore minimizing the cost functional over the class of adaptive strategies is expected to improve the results obtained when minimizing over the static class.    
In \cite{brigo18} this difference in the costs and in some cases risks was examined for two optimal trading frameworks: the discrete time Bertsimas and Lo model with an information signal and the continuous time Almgren and Chriss model that was studied by Gatheral and Schied in \cite{GatheralSchied}. In both frameworks, the difference between the transaction costs resulting from the optimal adaptive strategies and the corresponding optimal static strategies were negligible, except in cases where one took unrealistic parameter values for either the asset dynamics or the market impact function. One of the main questions which was left open in \cite{brigo18}, was whether there is any optimal trading framework in which the difference between the costs of adaptive vs static strategies will be considerable in a realistic setting. {\emph{The main goal of this paper is to point out one such trading framework. }}

We use the modelling framework from \cite{Lehalle-Neum18}, an optimal trading framework that incorporates \emph{signals} (i.e. short term price predictors) into optimal trading problems was established. It is important to notice that the purpose of this paper is not improving on the model of Lehalle and Neumann, but rather compare the static and dynamic optimality in this model as a fundamental case where the two classes of solutions may lead to quite different optimizers.  
As we mentioned earlier, usually optimal execution problems focus on the tradeoff between market impact and market risk. In the simplest models we discussed above there is no continuous signal related to price predictors in the dynamics.. However, in practice many traders and trading algorithms use short term price predictors. 
Most of such documented predictors relate to orderbook dynamics \cite{citeulike:12820703}.
An example of such signal is the order book imbalance signal, measuring the imbalance of the current liquidity in the limit order book.     

We will consider the following two types of market impact: instantaneous market impact and transient market impact with an exponential decay.  
In section \ref{sec-temp} we compare the optimal static strategy to the optimal adaptive strategy in the case where the market impact is instantaneous. In a further contribution of the paper, we  derive the static strategy in this setting though calculus of variations. Then, we show that there is a significant improvement in the expected revenues minus risk when the agent trades with the optimal adaptive strategy. 

In section \ref{sec-transient} we consider the transient market impact case. The optimal static strategy in this case was derived in \cite{Lehalle-Neum18}, however, finding the optimal adaptive strategy remains an open problem. 
We propose a strategy which uses the value of the signal a few times during the trading window. This strategy, even though not necessarily optimal, increases the revenue of the agent significantly.

\section{The instantaneous market impact case} \label{sec-temp}  
In this section we define a model which incorporates a Markovian signal into the optimal trading framework with instantaneous market impact.

We consider a filtered probability space $(\Omega,\mathcal F, (\mathcal F_{t}), P)$ satisfying the usual conditions, where $\mathcal F_{0}$ is trivial. Let $\wt W=\{\wt W_{t}\}_{t\geq 0}$ be a Brownian motion and $I=\{I_{t}\}_{t\geq 0}$ a homogeneous c\`adl\`ag Markov process satisfying,
\be \label{I-asmp} 
E_{\iota}\big[|I_{t}|] \leq C(T)(1+|\iota|), \quad \textrm{for all } \iota \in \re,  \ 0\leq t\leq T, 
\ee 
for some constant $C(T)>0$, where $T$ is the final execution time. Here $E_{\iota}$ represents expectation conditioned on $I_{0}=\iota$.

In our model $I$ represents a signal that is observed by the trader. 
We assume that the asset price process $P$, which is unaffected by trading transactions, is given by
\be \label{price} 
P_t = P_0 + \int_{0}^{t} I_s ds + \sigma_P \wt W_{t} , 
\ee  
hence the signal interacts with the price through the drift term, modeling the local trend of the price process. Here $\sigma_{P}$ is a positive constant modeling the price volatility. 

The rationale for having $I$ as drift in $P$ is the following. Suppose that $I$ is related to the 
order book imbalance \emph{Imb}. Such an imbalance measures the current liquidity in the limit order book according to the following formula by using the quantity of the best bid $Q_{B}$ and the best ask $Q_A$ of the order book, 
$$\mbox{Imb}(\tau)=\frac{Q_B(\tau) - Q_A(\tau)}{Q_B(\tau) + Q_A(\tau)},$$
where $Q_{B}$ and $Q_A$ are the quantity of limit orders at the best bid price and at the best ask price respectively. 
If $Imb>0$, we know that more participants want to buy than sell, and the price will move up. The opposite will tend to happen if $Imb$ is negative. This is the intuition on why $I$ is the correct drift for the price $P$. 

Let $\mathcal V$ denote the class of progressively measurable control processes $r=\{r_{t}\}_{t\geq0}$ for which 
$\int_{0}^{T}|r_{t}|dt<\infty$, $P$-a.s. 

If $x \geq 0$ denotes the initial amount of inventory,  we let 
\be \label{inv-eq}
X_{t}^{r}:=x-\int_{0}^{t}r_{s}ds. 
\ee
be the inventory trajectory with liquidation rate $r$; its marginal $X^{r}_{t}$ is the amount of inventory held by the trader at time $t$. We will often suppress the dependence of $X$ on $r$, to ease the notation. 
Note that  $r_t = - \dot{X}_t$, namely the trader's control is the trading speed.
The price at which orders are executed  is given by
\bn
S_{t} = P_{t}-\kappa r_{t}, \quad t\geq 0, 
\en
where $\kappa$ is a non-negative constant. This models the instantaneous linear market impact introduced in \cite{OPTEXECAC00}. We observe that the affected price is impacted by the trading speed $r$, which is typically positive. Hence for positive $\kappa$ the impacted price $S$ will be smaller than the ``mid'' price $P$.

Finally, the investor's cash $\mathcal C_{t}$ is defined as follows
\be \label{cash} 
d\mathcal C_{t} := -S_t dX^r_t = S_{t}r_{t}\,dt =(P_{t}-\kappa r_{t})r_{t}\,dt,  \quad \mathcal C_0 =0. 
\ee
with $\mathcal C_{0} =c$. Intuitively, $-S_t dX^r_t \approx S_{t+dt} (X^r_t - X^r_{t+dt})$ which is the revenue obtained from trading the inventory's portion $X^r_t - X^r_{t+dt}$ at the affected price $S_{t+dt}$ in the time interval $[t,t+dt]$.

The purpose of the execution would be, ideally, to complete the order by time $T$ and have zero remaining inventory, $X_T=0$. However, this is not always possible in practice. Therefore, as in Section 3 of \cite{Lehalle-Neum18}, we add a penalty function $-\varrho X^2_{T}$ for the remaining inventory at time $T$ that has not been executed. Here $\varrho$ is a positive constant which is used to adjust the weight of penalty. Another ingredient in our optimal execution problem is the risk aversion term, which reflects the risk associated with holding a position $X_t$ at time $t$. A natural candidate is the quadratic variation of the cash process, 
\[  \langle {\mathcal C} \rangle_t  = \sigma_P^2 \int_0^t X_u^2 du \]
which is similar to considering the variance of the cost. We can add a leverage parameter $\hat{\phi}$ that will allow us to specify the relative size of risk relative to cost $-\mathcal C$. This results in   
 $\phi \int_0^TX_t^2\,dt$, where 
 \be \label{phi-h} 
 \phi=\sigma_P^2 \hat{\phi},
 \ee
is a positive constant, see~\cite{AlmgrenSIFIN,Forsythetal,Tseetal} and the discussion in Section 1.2 of~\cite{SchiedFuel}. This term penalizes larger inventories. In absence of market impact, the optimal execution here would be liquidating the whole $X$ immediately. In presence of impact, however, this would lead to a very large speed $r$, leading to a very high cost term $\kappa r$. This tends to offset the low risk term, so that we end up with a compromise between keeping risk low and keeping impact low.  When the value of $\phi$ is high, risk is emphasized with respect to cost and the trading speed tends to be higher at the beginning of the execution, i.e. the execution becomes more urgent. Finally, we add the term $P_TX_T$ which is the final value of the remaining inventory.
The revenue-risk functional of the liquidation problem is 
    
\begin{equation}  \label{v-cost} 
E_{\iota,x,p}\Big[\mathcal C_{T}- \phi\int_{0}^{T}X^{2}_{s}ds+X_{T}(P_{T}-\varrho X_{T})\Big],
\end{equation} 
where  $E_{\iota,x,p}$ represents expectation conditioned on $I_{0}=\iota, X_{0}=x, P_{0}=p$.  

We first formulate the optimal adapted solution relying on \cite{Lehalle-Neum18}. 
Introduce the following functions 
\be \label{v-def} 
 \begin{aligned}  
 v_{2}(t) &= \sqrt{\kappa \phi} \frac{1+\zeta e^{2\beta(T-t)}}{1-\zeta e^{2\beta(T-t)}},   \\
 v_{1}(t,\iota)&=\int_{t}^{T} E[I_{s} \vert I_t = \iota] \exp \left(\frac{1}{\kappa}\int_{t}^{s}v_{2}(u)du\right)ds, \\
 v_{0}(t,\iota)  &=\frac{1}{4\kappa}\int_{t}^{T}E\big[v^{2}_{1}(s,I_{s}) \vert I_t = \iota\big]ds, 
 \end{aligned}
 \ee 
where the constants  $\zeta$ and $\beta$ are given by 
\be \label{constants}  
\zeta = \frac{\varrho+ \sqrt{\kappa \phi}}{\varrho- \sqrt{\kappa \phi}}, \qquad  \beta = \sqrt{\frac{ \phi}{\kappa}}.
\ee
If $\varrho \not = \sqrt{\kappa \phi}$, then the maximizer of the revenue functional in (\ref{v-cost}) exists, is unique and given by    
\be \label{r-signal-stoch} 
r^{*}_{t}=  -\frac{1}{2\kappa}\Big(2v_{2}(t)X_{t}+\int_{t}^{T}e^{\frac{1}{\kappa}\int_{t}^{s}v_{2}(u)du} E[I_{s} \vert I_t ]ds\Big), \quad 0\leq t\leq T,
\ee
where, for $s>t$, $E[I_{s} \vert I_t ]$ is the expected value of $I_s$ given $I_t$. It is such reaction to the signal $I_t$ that accounts for the adaptiveness of $r^{*}_t$. 
The optimal revenue is given by $c-xp+ v_{0}(0,\iota) + xv_{1}(0,\iota)+x^{2}v_{2}(0)$.


We now focus on the case where $I$ follows an Ornstein-Uhlenbeck process, 
\be
\begin{aligned} \label{I-OU} 
dI_{t}&= -\gamma I_{t}\, dt +\sigma \, dW_{t},    \quad t\geq 0, \\
I_{0}&=\iota, 
\end{aligned} 
\ee
where $W$ is a standard Brownian motion independent of $\widetilde{W}$ and  $\gamma, \sigma>0$ are constants.  The choice of a mean reverting model for the imbalance is based on the following. If $Imb>0$, more participants want to buy, but new participants who are keen to buy may  post a limit order at a higher price than current best bid, in order to avoid the long queue. Price will then go up and  imbalance evens out. For more discussion see \cite{Lehalle-Neum18}. The parameter $\gamma$, if positive, is the speed of mean reversion to zero for the signal starting at $I_0$. The parameter $\sigma$ is the signal absolute volatility.
Then, $r^*$ has the form 
\bn
r^{*}_{t}=  -\frac{1}{\kappa}v_{2}(t)X_{t}+-\frac{1}{2\kappa} I_{t}\int_{t}^{T}\exp \left(-\gamma (s-t)+ \frac{1}{\kappa}\int_{t}^{s}v_{2}(u)du\right)ds, \quad 0\leq t\leq T.
\en
\begin{remark} 
One can impose a constraint on the admissible strategies to terminate without any inventory, that is to have $X_T=0$. This constraint is often called a ``fuel constraint'' as the strategy is forced to terminate without any ``fuel''. 
In our setting we could heuristically impose a fuel constraint on the strategy that maximizes (\ref{v-cost}) by using the asymptotics of $r_{t}^{*}$ when $\varrho \rr \infty$. In this case $\zeta \rr 1$ and the limiting trading speed, which we denote by $r_{t}^{f}$, is 
\be \label{dynamic-fuel} 
r^{f}_{t}=  -\frac{1}{2\kappa}\Big(2\bar v_{2}(t)X_{t}+I_{t}\int_{t}^{T}e^{-\gamma (s-t)+ \frac{1}{\kappa}\int_{t}^{s}\bar v_{2}(u)du}ds\Big), \quad 0\leq t\leq T,
\ee
where
\bn
\bar v_{2}(t) &=& \sqrt{\kappa \phi} \frac{1+ e^{2\beta(T-t)}}{1- e^{2\beta(T-t)}}.
\en
\end{remark} 
We note that the optimal solution $r^*$ does not depend explicitly on the price $S$ but is adaptive only through the signal $I$. Furthermore, in cases where the drift $I$ of the price $P$ is deterministic (for example if $\sigma =0$ in the Ornstein-Uhlenbeck process \eqref{I-OU}) one sees immediately that the quantities $r^*$ and $r^f$ above becomes static. This leads us to suspect that the optimal dynamic solutions collapse to static in cases where the drift $I$ is deterministic. We need however to prove this rigorously. 

In order to prove this claim, we define $d \bar P_t = \bar I(t) dt + \sigma_P d\widetilde{W}_t$ where $t \mapsto \bar I(t)$ is a continuous deterministic function.  
We also define the investor's cash $\bar{\mathcal C}_{t}$, similarly to (\ref{cash}),  
$$
d \bar{\mathcal C}_{t} :=  (\bar P_{t}-\kappa r_{t})r_{t}\,dt, \quad \bar{\mathcal C}_0=0. 
$$

We consider the following value function, which corresponds to the cost functional (\ref{v-cost}),  
\begin{equation}  \label{v-cost-det} 
 V(x,p) = \sup_{r\in \mathcal V}E_{x,p}\Big[\bar{\mathcal C}_{T}- \phi\int_{0}^{T}X^{2}_{s}ds+X_{T}(\bar P_{T}-\varrho X_{T})\Big].
\end{equation} 
Here $E_{x,p}$ represents expectation conditional on $X_{0}=x, \bar P_{0}=p$.  

Recall that $v_2(\cdot)$ was defined in (\ref{v-def}). Before we state our next result we define the following functions,  
\be
\begin{aligned} \label{bar_v} 
\bar v_1(t) &= \int_t^Te^{\frac{1}{\kappa} \int_t^u v_2(s)ds}\bar I(u)du,   \\
 \bar v_{0}(t)  &=\frac{1}{4\kappa}\int_{t}^{T}\bar v^{2}_{1}(s)ds. 
 \end{aligned}
\ee 
 In the following proposition we prove that when the signal is deterministic then the optimal trading speed must also be deterministic. 
\begin{proposition} \label{prop-stat-drift} 
The value function (\ref{v-cost-det}) is given by
\be \label{v-eq} 
V(t,p,x) = px+\bar v_0(t) +x\bar v_1(t)+x^2 v_2(t). 
\ee
Moreover, the unique optimal trading speed $r^* \in \mathcal V$, is  
\be \label{r-eq} 
r^*(t) =- \frac{1}{2\kappa}\big(\bar v_1(t) + 2X^*_tv_2(t)\big),
\ee
where $X^*_t = x+\int_0^tr^*_sds$. 
\end{proposition}
The proof of Proposition \ref{prop-stat-drift} is given in the Appendix.  

\begin{remark} 
In Proposition \ref{prop-stat-drift} we proved that when the signal $I(t)$ is deterministic, the optimal trading speed $r^*$ over the class of adapted admissible strategies $\mathcal V$ turns out to be deterministic.  The proof of the dynamic case involved the solution of a system of second order PDEs (see Eqs (5.15) to (5.17) in \cite{Lehalle-Neum18}) while the solution in the deterministic case only involved first order ODEs (see Eqs. \eqref{ap-1}-\eqref{ap-3}). The reason for this is that $I$ is no longer a Markov process and its generator does not appear in the system of equations, where $I$ appears as a time-varying coefficient in the HJB equation \eqref{a-hjb1}. The static and dynamic approaches can be reconciled in the spirit of \cite{muhle-karbe}.

\end{remark}

We now solve the static optimization under a fuel constraint. If $x$ denotes the quantity of asset to be liquidated, this means that the admissible strategies are those in the set 
$$\mathcal V_{S}(x) = \Big\{r : r \ \mbox{is deterministic}, \  \int_{0}^{T} \vert r_s \vert ds < \infty  \textrm{ and } X^r_0-X^r_T =\int_{0}^{T}r_{s}ds=x\Big\}. $$
Notice that $\mathcal V_{S}$ is a subset of $\mathcal V$. As a consequence of such choice, the revenues functional will no longer have the penalisation on the inventory left after trading, and it will be defined as  
\begin{equation}  \label{stat-cost} 
E_{\iota,c,x,p}\Big[\mathcal C_{T}-\phi\int_{t}^{T}X^{2}_{s}ds\Big].
\end{equation}

In the following Theorem, we derive a necessary and sufficient condition to the maximiser of (\ref{stat-cost}) over the class of admissible strategies $\mathcal V_{S}(x)$. 
\begin{theorem}  \label{thm-opt-stat}
$r^{*} $ maximizes the revenue functional (\ref{stat-cost}) over $\mathcal V_{S}(x)$, if and only if there exists a constant $\lambda$ such that $r^{*}$ solves
\begin{equation}\label{cond-opt} 
2kr^{*}_t +2\phi\int_{0}^{t}X^{*}_{s}ds -\int_{0}^{t}E_{\iota}[I_{s}]ds  =\lambda, \quad \textrm{for all } 0\leq t\leq T, 
\end{equation}
where $X^{*}_{t} = x- \int_{0}^{t}r^{*}_{s}\,ds$. 
\end{theorem} 

Recall that $\beta$ was defined in (\ref{constants}). From Theorem \ref{thm-opt-stat} we can easily deduce the following corollary. 
\begin{corollary}  \label{corr-stat} 
Assume that $I$ follows an OU-process as in (\ref{I-OU}). Then, the optimal static inventory $X^*:=X^{r^*}$  is given by 
\be \label{stat-min}
X^{*}_{t}= x \psi (t) + \varphi (t), 
\ee
where $\psi(t) = \frac{\sinh(\beta(T-t))}{\sinh(\beta T)}$ and 
\be 
\varphi(t)=\frac{I_{0}}{2\kappa(\beta^{2}-\gamma^{2})}\Big(1-\frac{e^{-\gamma(T-t)} \sinh(\beta t) +e^{\gamma t} \sinh(\beta(T-t)}{\sinh(\beta T)}\Big).
\ee
\end{corollary}  
In Figure \ref{inv-pic2} we present the optimal static inventory $X^{*}$ in (\ref{stat-min}) for the parameters: $\gamma=0.1$, $\sigma =0.1$, $T=10$, $\kappa=0.5$, $\hat \phi=0.1$, $X_{0}=10$, $\sigma_P=1$, and therefore by (\ref{phi-h}), $\phi =0.1$. The influence of the initial value of the signal on the optimal strategy is demonstrated for $I_0 = 0.5$, $I_0 = 0$ and $I_0 = -0.5$. Since $I$ represents the local trend of the price $P$, we are assuming quite significant trends of $50\%$ and $-50\%$. Typical values of the signal which may initiate trading for static strategies appear in Fig. 4.2 and Fig 4.6 top left in \cite{Lehalle-Neum18} and $50\%$ is in this range. In later examples we will adopt $\pm20\%$. In Figure \ref{inv-pic2-sig} we present the optimal static inventory $X^{*}$ in (\ref{stat-min}) for the same parameters as in Figure \ref{inv-pic2}, only now we set: $I_0=0.2$ and we show the influence of the asset volatility on the optimal strategy for $\sigma_P = 1$, $\sigma_P = 5$ and $\sigma_P = 10$. We can see that large volatilities bring down the inventory schedule faster. This is because, with large volatility, the risk component of the criterion becomes more important compared with the revenues part. 

The reminder of this section is dedicated to a comparison between the signal adaptive strategy $r^f$ in (\ref{dynamic-fuel}) and the optimal static strategy $X^*$ from (\ref{stat-min}), and the comparison of their corresponding revenues.  In Figure \ref{inv-pic} (blue region) we simulate $1000$ trajectories of the inventory $X^{r^f}$ resulting from $r^{f}$ . In the black curve we present the optimal static inventory from (\ref{stat-min}). For the signal process $I$ parameters and the execution problem impact and boundary conditions we assume the following values:
\begin{equation}\label{eq:paramOU}
\gamma=0.1,\ \sigma =0.1,\ \ I_0=0.2, \ T=10,\ \kappa=0.5,\ \hat\phi=0.1,\  \sigma_P=1, \ X_{0}=10. 
\end{equation}
The parameters of the model are similar to the parameters of Figure \ref{inv-pic2} with the addition of $I_0=0.2$. We notice that even though the strategies start and end with the same innovatory values, the changes in the trading speed during $(0,T)$ can be substantial. 

 In Figure \ref{rev-pic} (left) we compare the revenues resulting from the optimal static strategy (\ref{stat-min}) in blue, and the signal adaptive strategy (\ref{dynamic-fuel}) in orange. The revenues are plotted for different values of trading windows $T$ from $5$ to $50$. We observe that as the trading window increases, the difference in the expected revenues of the strategies increases drastically. In Figure \ref{rev-pic}  (right) we compare the revenues for different values of signal volatility $\sigma$. The model parameters (except form $\sigma$) are similar to the left plot. We observe that a signal with a large volatility will create a major difference between the revenues of the static and adaptive strategies. 
\begin{figure}
\centering 
\includegraphics[width=10cm]{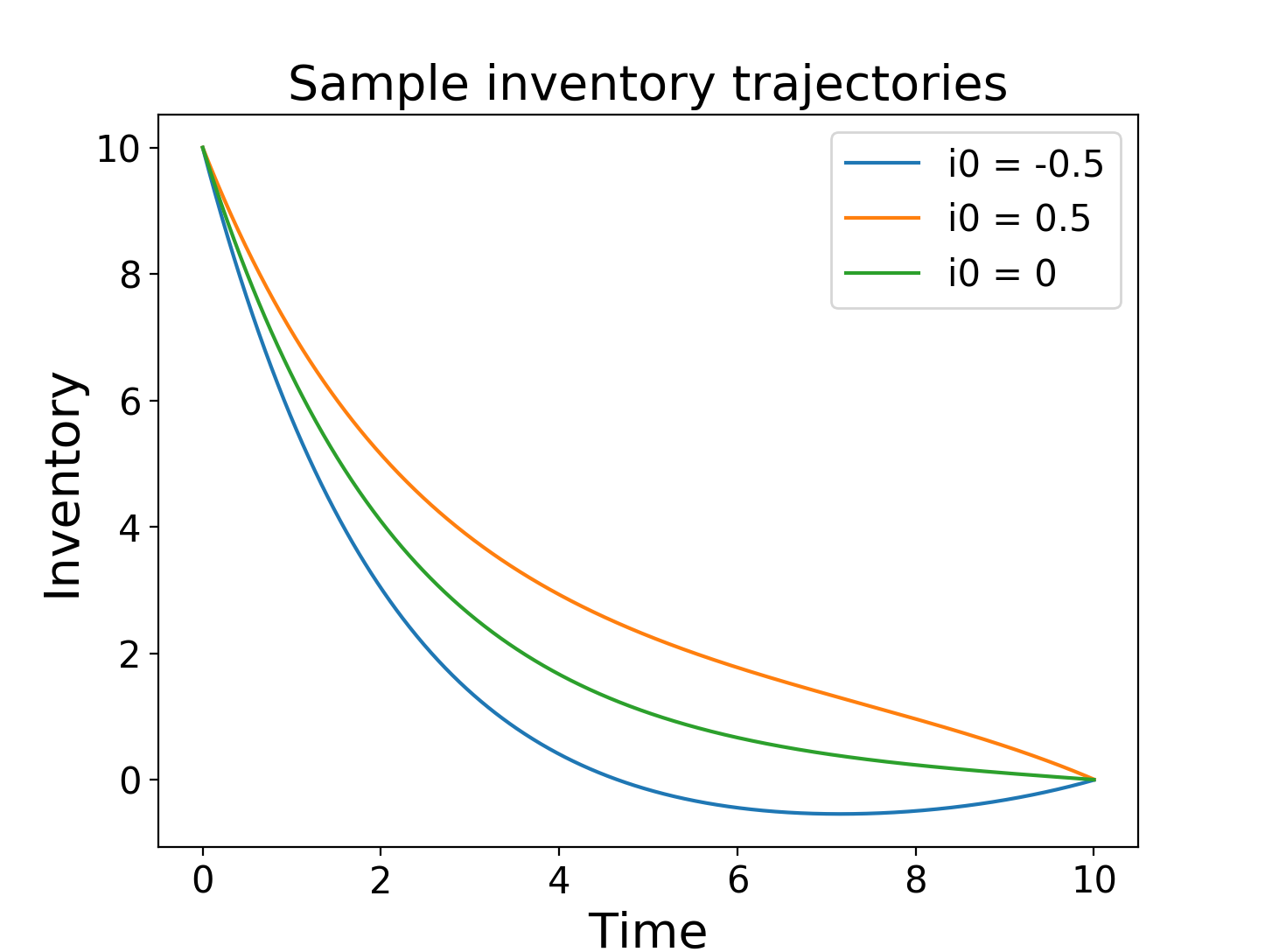} 
\caption{Plot of the optimal static inventory $X^{*}$ in (\ref{stat-min}), for the parameters in \eqref{eq:paramOU} except for $I_0$. The optimal static strategy is presented for different initial values of the signal: $I_0 = 0.5$ (orange), $I_0 = 0$ (green) and $I_0 = -0.5$ (blue).}\label{inv-pic2}
\end{figure}   

\begin{figure}
\centering 
\includegraphics[width=10cm]{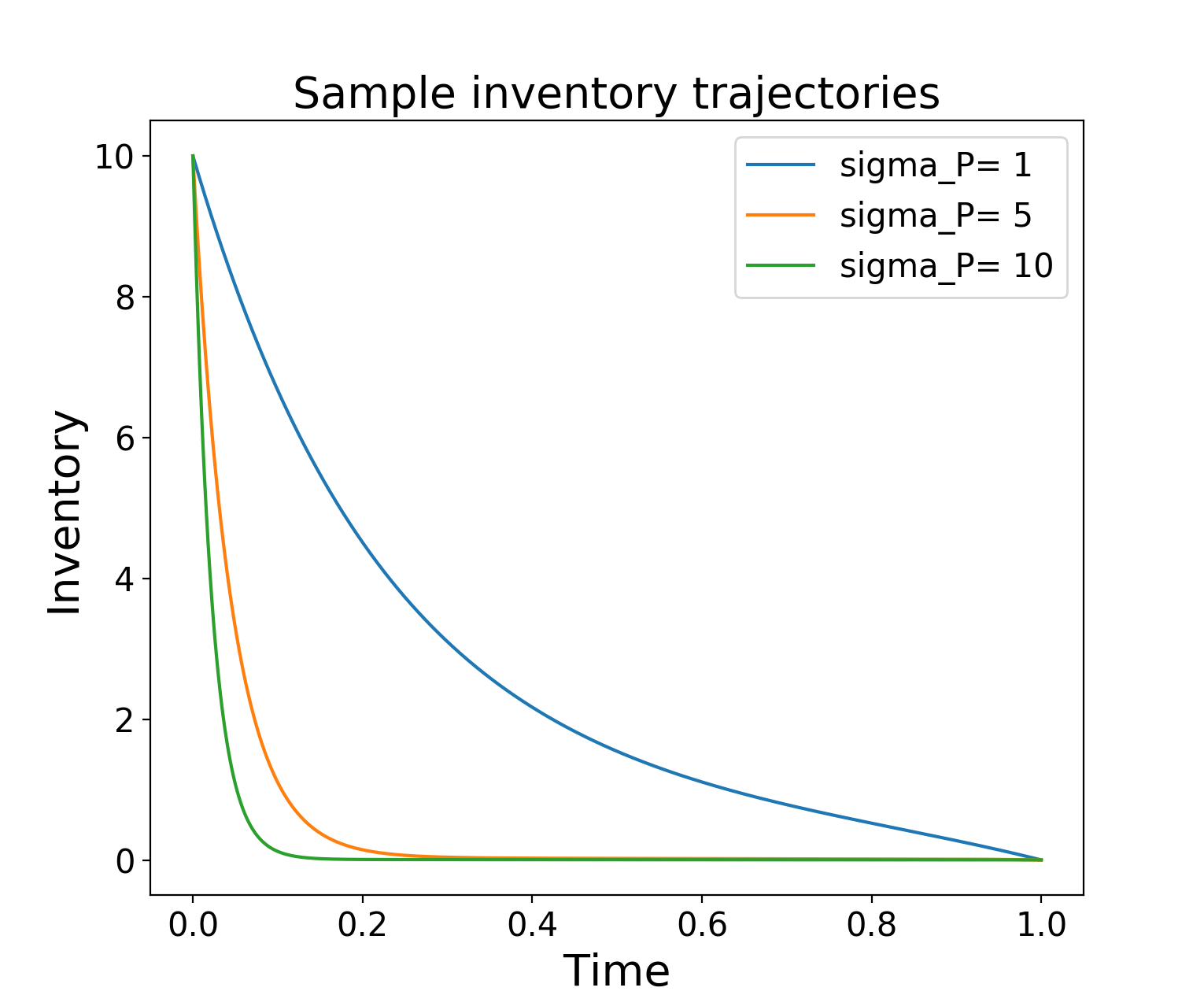} 
\caption{Plot of the optimal static inventory $X^{*}$ in (\ref{stat-min}), for the parameters in \eqref{eq:paramOU} except for $\sigma_P$. The optimal static strategy is presented for different values of the volatility: $\sigma_P = 1$  (blue), $\sigma_P= 5$ (orange) and $\sigma_P = 10$ (green).}\label{inv-pic2}
\end{figure}  \label{inv-pic2-sig}

\begin{figure}
\centering 
\includegraphics[width=10cm]{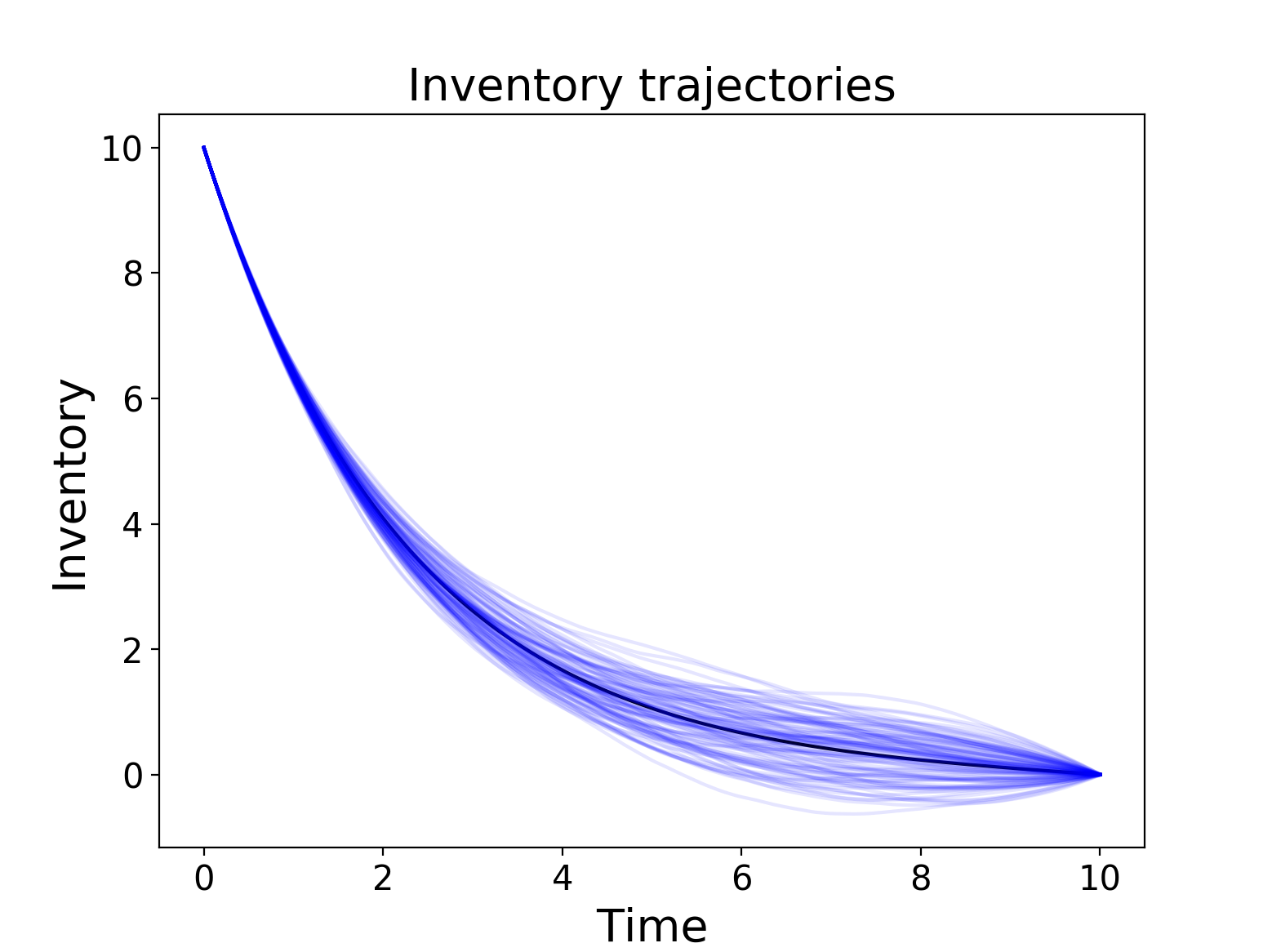} 
\caption{Simulation of the inventory $X^{r^f}$ resulting from the signal adapted trading speed $r^{f}$in (\ref{dynamic-fuel}). The blue region is a plot of $1000$ such trajectories of $X^{r^f}$. In the black curve we present the optimal static inventory (\ref{stat-min}). The parameters of the model are as in \eqref{eq:paramOU}. }\label{inv-pic} 
\end{figure}

 \begin{figure}
\centering 
\includegraphics[width=7cm, height =6.5cm]{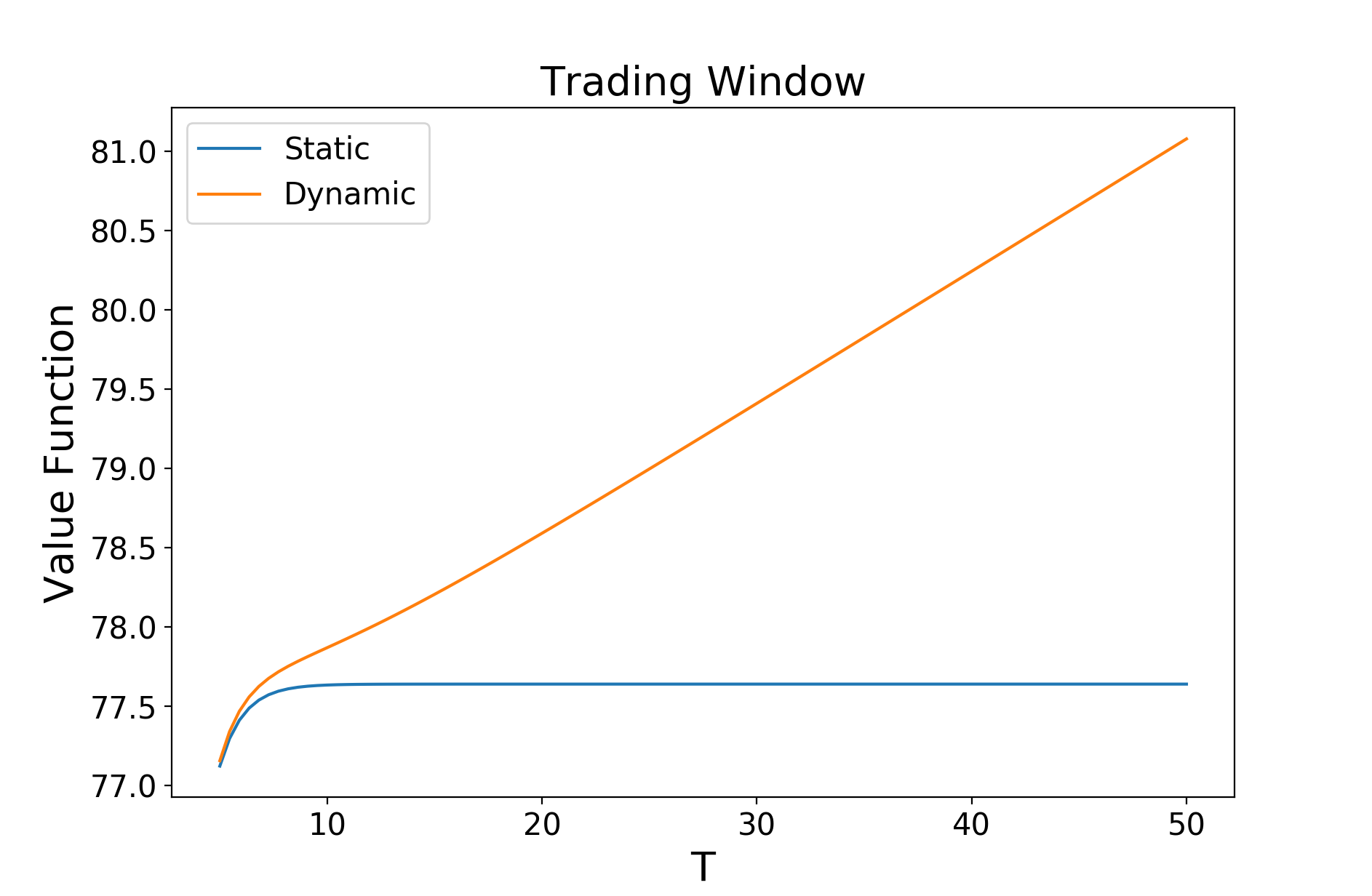}  \
\includegraphics[width=7cm, height =6.5cm]{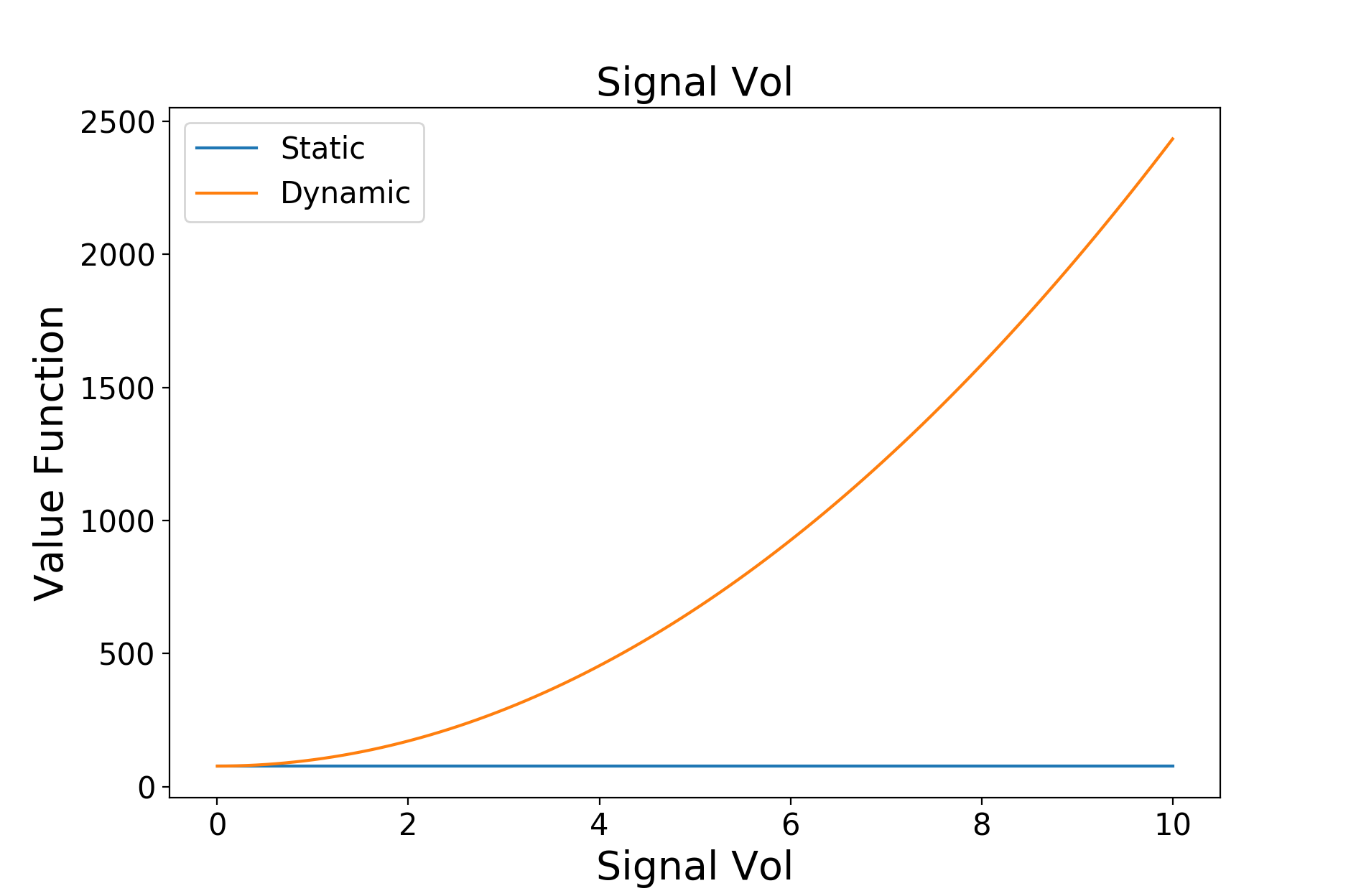}   
\caption{Left: comparison of the revenues resulting from the optimal static strategy (\ref{stat-min}) in blue, and the signal adaptive strategy (\ref{dynamic-fuel}) in orange. The revenues are plotted for different values of trading windows $T$. The parameters of the model are as in \eqref{eq:paramOU} plus $P_0=10$. Right: comparison of the revenues for different values of signal volatility $\sigma$. The model parameters (except form $\sigma$) are similar to the previous plot. } \label{rev-pic} 
\end{figure}

\newpage

\section{The transient market impact case} \label{sec-transient} 
In this section consider the case where the market impact is exponentially decaying as in the Obizhaeva and Wang model  \cite{Ob-Wan2005}. The actual price process in this model is given by 
\be \label{market-imp}
S_{t} = P_{t}+\kappa\rho\int_{\{s<t\}} e^{-\rho(t-s)}dX_{s}, \quad t\geq 0, 
\ee 
where $P$ and $I$ are given as in (\ref{price}) and (\ref{I-OU}), receptively, and $\kappa, \, \rho$ are positive constants. 
In this context we say that the inventory $X$ is an admissible strategy, if it satisfies:
\begin{itemize} 
\item [\textbf{(i)}]  $t\longrightarrow X_{t}$ is left--continuous and adapted. 
\item [\textbf{(ii)}] $t\longrightarrow X_{t}$ has  $\mathbb{P}$-a.s. bounded total variation. 
\item [\textbf{(iii)}] $X_{0}=x$ and $X_{t}=0$, $\mathbb{P}$-a.s. for all $t>T$. 
\end{itemize} 
For the sake of readability we will assume that the risk-aversion constant $\phi=0$. It was shown in Section 2.1 of \cite{Lehalle-Neum18} that the revenue functional which corresponds to an admissible strategy $X$ is given by 
\be \label{costs-trans} 
P_0x-E\Big[\int \int_{0}^{t}I_{s}\,ds\,dX_{t}+\frac{\kappa \rho}{2} \int \int \rho e^{- |t-s|}dX_{s}dX_{t}\Big]. 
\ee
The class of static strategies in this case is defined as follows,   
\begin{equation*} 
\Xi_S(x) = \{ X | \textrm{ deterministic admissible strategy with } X_{0}=x \ \textrm{and support in  }  [0,T]  \}. 
\end{equation*}
In Corollary 2.7 of \cite{Lehalle-Neum18} the unique static strategy $X^*$ which maximises the revenue functional (\ref{costs-trans}) was derived, 
 \be \label{opt-spec}
X^{*}_{t}=(1-b_0(t))\cdot x + {\iota \over 2\kappa\rho^2\gamma}\left\{ \frac{\rho^2-\gamma^2}{\gamma} \cdot b_1(t) - (\rho + \gamma) \cdot b_2(t) - (\rho + \gamma) \cdot b_3(t)\right\},
 \ee
where
  \bn
b_0(t) &=& {\one_{\{t>0\}}+\one_{\{t>T\}}+ \rho t\over 2 + \rho T},\\
b_1(t) &=&1 - e^{-\gamma t} -b_0(t)(1-e^{-\gamma T}),\\
b_2(t) &=&\one_{\{t>T\}}+ \rho t - b_0(t)(1+\rho T),\\ 
b_3(t) &=&(b_0(t) - \one_{\{t>T\}}) e^{-\gamma T}.
 \en 
The optimal adaptive strategy for this model is an open problem (see Remark 2.9 in \cite{Lehalle-Neum18}). 
Note that $X^*_t$ has jumps at $t=0$ and $t=T$ and is continuous for $0<t<T$. Moreover, $X_t$ is a function of the initial signal value $\iota$, initial inventory $x$, initial time (which is set to $0$ in (\ref{opt-spec})) and the terminal time $T$. In what follows we will write $X_t^*(I_s, x, s,T)$, for the optimal static strategy which starts at time $0\leq s \leq T$ when the signal value is $I_s$, the inventory held the trader at the initial time $s$ is $x$, and it terminates at time $T$ (with $X_{T}=0$). 

We will now propose a dynamic strategy which improves the results of the optimal static strategy $X^{*}$. This new strategy $\wt X^{(n)}$, allows the agent to update the trading strategy at $n-1$ intermediate times according to the new information available at these times. To formalise this we choose $n\geq 1$ and define a grid on $[0,T]$ such that $t_k = \frac{kT}{n}$, $k=0,...,n$. We also define 
\be \label{trans-dynamic} 
\wt X_t^{(n)} = \begin{cases}
X_0, &\text{ if } t=0,\\
X^*_t(I_{t_{k-1}}, \wt X^{(n)}_{t_{k-1}}, t_{k-1},T), &\text{ if } t_{k-1} < t\leq t_k, \quad  k=1,...,n. 
\end{cases}
\ee
Note that $X^*= \wt X^{(1)}$. 

\begin{remark} 
We remark at this point that is not a-priori trivial that the revenue which is associated with $\wt X_t^{(n)}$ $n\geq2$ is larger than the revenue of $X^*$. Since the market impact is transient and does not vanish immediately, a trader who updates his strategy at time $T/2$ for example according to $X^*_t(I_{T/2}, \wt X^{(2)}_{T/2}, T/2,T)$, does not take into account the market impact which is caused by his strategy on the interval $[0,T/2]$, hence his strategy may be suboptimal (see Remark 2.9 in \cite{Lehalle-Neum18} for detailed discussion). 

\end{remark} 
In Figure \ref{trans-pic} we compare $\wt X^{(n)}$ with the optimal static strategy $X^*$. On the left panel, in the blue curves, we plot 50 trajectories of $\wt X^{(2)}$ where the update takes place at $t=5$. The black curve presents the optimal static strategy $X^*$ from (\ref{opt-spec}). One can observe that $\wt X^{(2)}$ has an additional jump at $T/2$ which is caused by the update of the strategy. 
On the right panel we show the results of Monte-Carlo simulations for the revenue functional (\ref{costs-trans}) which corresponds to $\wt X_t^{(n)}$, for $n=1$ (blue), n=2 (orange) and $n=3$ (green). Note that the case where $n=1$ is the static case. The graph shows the convergence of the expected revenue ($y$-axis) as a function of the number of trajectories $N$ ($x$-axis) in the simulation. We observe that an increasing  number of signals updates during the trading window improves the results of the execution, as the revenue functional increases. 
The parameters in both graphs are $\gamma=0.1, \sigma =0.1, I_0=0.2, T=10, \rho=1, \kappa=0.5, X_{0}=10$ and $P_0=10$.

 \begin{figure}
 \includegraphics[width=8cm, height =5cm, trim={0 0 0 2.25cm},clip]{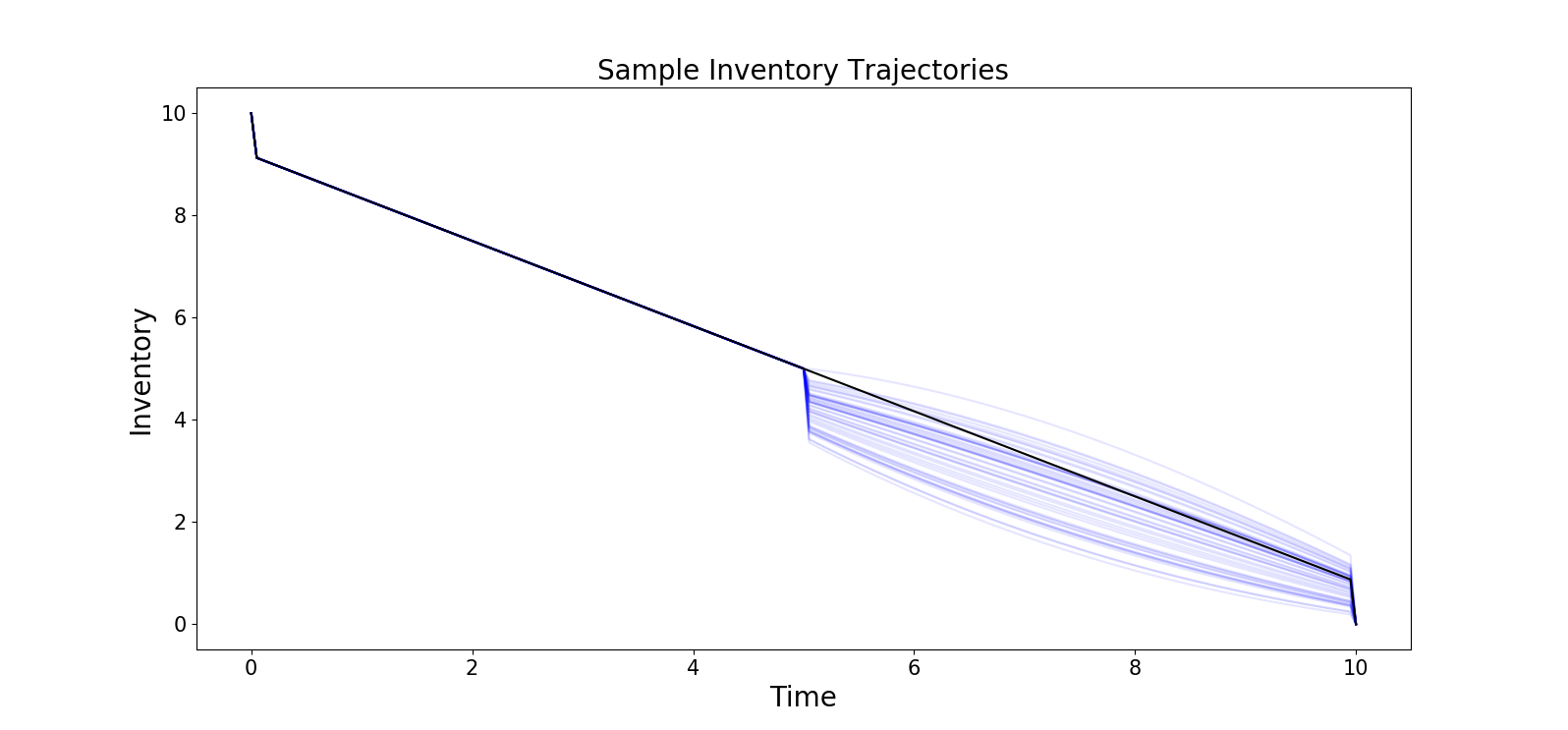}   
\includegraphics[width=8cm, height =5cm, trim={0 0 0 2.25cm},clip]{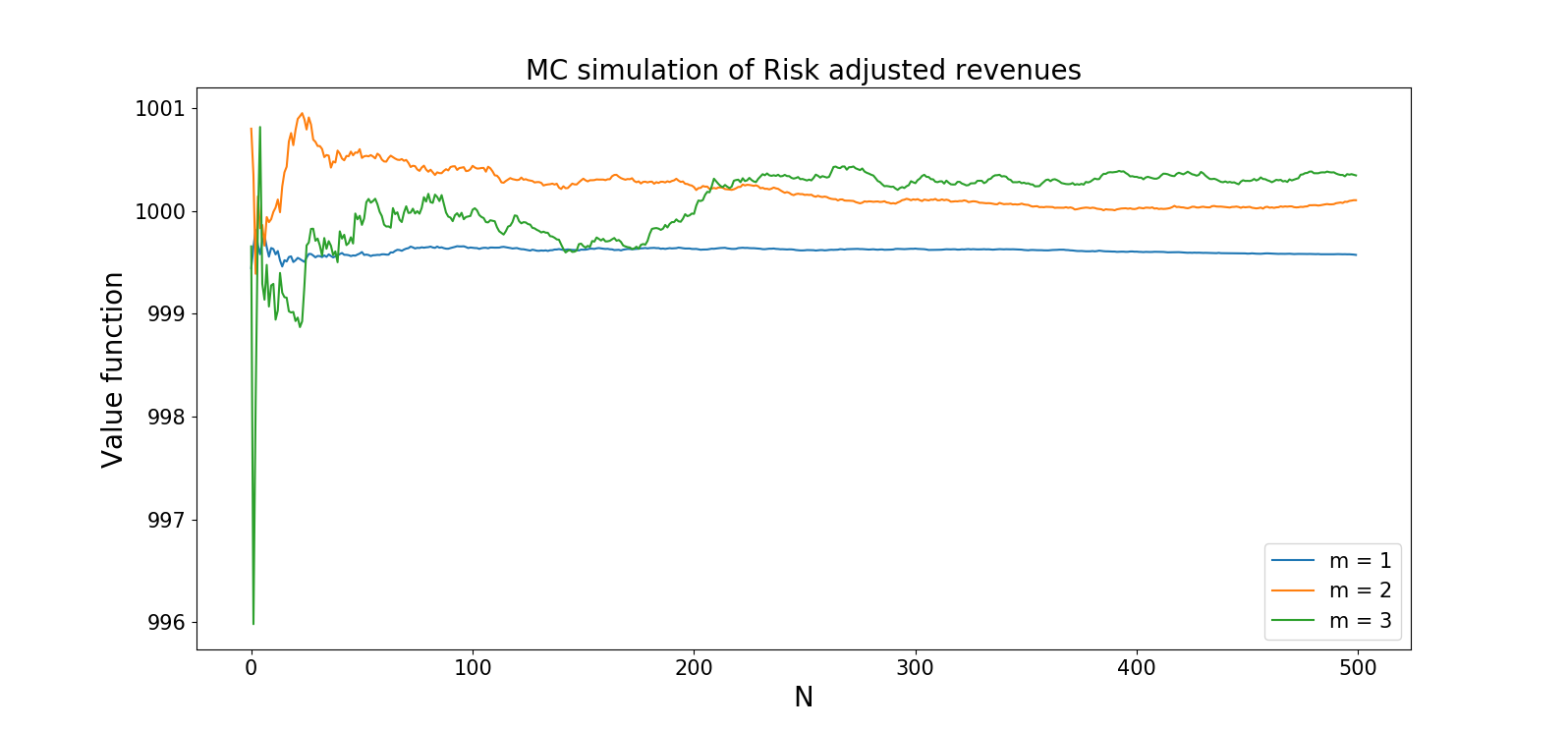}   \ 
\caption{Left: simulation of 50 trajectories of $\wt X^{(2)}$ from (\ref{trans-dynamic}), where the update takes place at $t=5$ (blue curves). The black curve presents the optimal static strategy $X^*$ from (\ref{opt-spec}). 
 Right: Monte-Carlo simulations of the revenue functional (\ref{costs-trans}) which corresponds to $\wt X_t^{(n)}$, for $n=1$ (blue), n=2 (orange) and $n=3$ (green). The parameters in both graphs are $\gamma=0.1, \sigma =0.1, I_0=0.2, T=10, \rho=1, \kappa=0.5, X_{0}=10$ and $P_0=10$.}
\end{figure}   \label{trans-pic}

\section{Conclusions and further research}\label{sec-conclusion}
In this work we investigated trade execution models in which the optimal adaptive strategy differs significantly from the static one. Previous results of Brigo and Piat \cite{brigo18} considered the benchmark models of Bertsimas and Lo with information signal \cite{BLA98} and of Gatheral and Schied \cite{GatheralSchied} after Almgren and Chriss \cite{OPTEXECAC00}. Under these models the improvement in optimality expected from adaptive strategies was found to be minimal, at least for reasonable values of the model parameters. To find models where the improvement is substantial,  we considered the trading framework proposed and studied by Lehalle and Neuman \cite{Lehalle-Neum18}. Such a framework incorporates the usage of price predictors in optimal trade execution, reconciling the academic literature  with traders' practice.   We found that within Lehalle and Neuman's model the improvement  can indeed be appreciated  with realistic values of the model parameters. Therefore, our conclusion is that  switching from static to adaptive strategies does pay off, but this is captured only by models that are sophisticated enough to incorporate some market practice. In future research the static-adaptive comparison could be extended to broader classes of models.

%

\bibliographystyle{plain}
\printindex

\begin{appendix}

\section{Proofs}
\paragraph{Proof of Proposition \ref{prop-stat-drift}.} 
Note that the HJB equation associated with \eqref{v-cost-det} is given by, 
\begin{equation}\label{a-hjb1} 
\partial_t V +\bar I(t)  \partial_p V + \frac{\sigma_P^2}{2} \partial_{pp} V - \phi x^2 + \sup_r \left\{ - r \partial_x V + p r - \kappa r^2\right\} =0, 
\end{equation}
with the terminal condition $V(T,x,p) = x(p-\varrho x)$. Plugging in the ansatz $V(t,x,p)= xp+v(t,x)$, we get that $v$ satisfies 
\[ 
\partial_t v + x\bar I(t) - \phi x^2 + \sup_r \left\{ - r \partial_x v  - \kappa r^2\right\} =0. 
\]
Optimising over $r$ we get
\be \label{r-star} 
 r^* = -\frac{\partial_x v }{2 \kappa},
 \ee
 and it follows that we need to solve the following PDE:
\begin{equation}\label{eq:HJBdet} \partial_t  v + x\bar I(t)  - \phi x^2 + \frac{1}{4 \kappa}\partial_x v^2 =0,
\end{equation}
with the terminal condition $v(T,x) = -\varrho x^2.$
By assuming that $v(t,x) = v_0(t) +xv_1(t)+x^2 v_2(t)$ and comparing similar powers of $x$, we get the following system of equations
 \bq
\label{v0}\partial_{t} v_{0} + \frac{1}{4\kappa}v_{1}^{2} &=&0,  \label{ap-1}\\   
\label{v1}\partial_{t} v_{1} + \frac{1}{\kappa}v_{2}v_{1}+ \bar I(t) &=&0,  \label{ap-2}\\
\label{v2} \partial_{t} v_{2} +\frac{1}{\kappa}v_{2}^{2} -\phi&=&0, \label{ap-3}
\eq
with the terminal conditions
$$
v_{0}(T)=0, \quad v_{1}(T) =0, \quad v_{2}(T)= -\varrho. 
$$
Note that (\ref{ap-3}) is the Riccati equation and that (\ref{ap-2}) is solved by an integration factor, so we get (\ref{bar_v}) and
(\ref{v-eq}). Equation (\ref{r-eq}) follows from (\ref{r-star}) and (\ref{v-eq}). 

The fact that $V$ is the value function (\ref{v-cost-det}) follows from Theorem 3.5.2 of \cite{pham}. The uniqueness of the optimal strategy follows by the same argument in Proposition 3.2 of \cite{Lehalle-Neum18}. 
 \qed

\paragraph{Proof of Theorem \ref{thm-opt-stat}} 
We will first prove the uniqueness of the optimal strategy. Let $x> 0$. For any $r\in \mathcal V_{S}(x)$ define 
\be \label{gen-cost} 
C(r):=C_{1}(r)+C_{2}(r) - K(r), 
\ee
where
\bd
C_{1}(r) =\kappa \int_{0}^{T} r_{s}^{2}\,ds,  \quad C_{2}(r) = \phi\int_{0}^{T}X_{t}^{2}dt, \quad K(r)=\int_{0}^{T} \int_{0}^{t}  E_{\iota}[I_{s}]\,ds\,r_{t}dt .
\ed
Note that $C(x)$ is the revenue functional in (\ref{stat-cost}) with a minus sign. 
From the fuel constraint and since $x>0$ we have  
\be \label{pos-c1} 
C_{1}(r)> 0, \quad C_{2}(r) > 0. 
\ee
Let $r,v \in \mathcal V_{S}(x)$. We define the following cross functionals, 
\be
C_{1}(r,v) = \kappa \int_{0}^{T} r_{s}v_{s}\,ds, \quad C_{2}(r,v)=\phi \int_{0}^{T}\int_{0}^{t}r_{s}v_{s}ds\,dt.
\ee 
Note that 
\bn
C_{i}(r,v)  = C_{i}(v,r), \quad \textrm{for } i=1,2, 
\en
and
\be \label{cx-y}
C_{i}( v- r) =C_{i}(v)+C_{i}(r) -2C_{i}(v,r), \quad \textrm{ for  } \ i=1,2. 
\ee

We now can repeat the same steps as in the proof of Theorem 2.3 in \cite{Lehalle-Neum18} and argue that $C(\cdot)$ is strictly convex to obtain existance of at most one minimizer to $C(\cdot)$ in $\mathcal V_{S}(x)$.

We now show that condition  (\ref{cond-opt}) is sufficient for optimality.  
Assuming that  $r^{*}\in \mathcal  V_{s}(x)$ satisfy (\ref{cond-opt}), we will show that $r^{*}$ minimizes $C(\cdot)$. Let $ r $ be any other strategy in $\mathcal V_{s}(x)$.  Define $v= r -r^{*}$ and note that from the fuel constraint it follows that $X^v_0=X_0^v=0$. 
We have
\bn
C(r)& = &C(v+r^*) \\
&=& C_1(r^*) +C_1(v) + C_1(r^*) +C_1(v) + 2C_1(r^*,v) 
 \\
&&\quad+C_2(r^*) +C_2(v) + 2C_2(r^*,v) \\
&&\quad  -K(r^*)-K(v)  \\
&=&C(r^*)+ C_1(v)+ C_2(v) -K(v) + 2C_1(r^*,v) +2C_2(r^*,v).
\en
Since $C_i(\cdot)\geq 0$, $i=1,2$, it follows that in order to prove the optimality of $r^*$ we need to show that 
$$
\ell(r^*,v):= 2C_1(r^*,v) +2C_2(r^*,v)-K(v)  \geq 0. 
$$
Use (\ref{cond-opt}) to get  
\bd\label{g1}
\aligned 
\ell(r^*,v)&=2\kappa \int_0^T r_t^*v_t dt + 2\phi\int_0^TX_t^vX_t^r{^*}dt - \int_0^T\int_0^tE_\iota[I_s]ds \,v_tdt  \\ 
&=\lam\int_0^Tv_tdt - 2\phi\int_0^T \int_0^tX_s^{r^*}ds\,v_tdt+ 2\phi\int_0^TX_t^vX_t^r{^*}dt. 
\endaligned 
\ed
From integration by parts we have 
$$
 \int_0^T \int_0^tX_s^{r^*}ds\,v_tdt- \int_0^TX_t^vX_t^r{^*}dt =0. 
$$
From the fuel constraint it follows that $\int_0^Tv_tdt =0$, and therefore $\ell(r^*,v) =0$. 
\qed 

\paragraph{Proof of Corollary \ref{corr-stat}}
In this case we have $E_{\iota}[I_{t}] = \iota e^{-\lam t}$. Assume a twice differentiable $r_{t}$ and differentiate both sides of (\ref{cond-opt}) to get
\begin{equation} \label{ode} 
-2k\ddot X_t+ 2\phi X_{s}-\iota e^{-\lam t} =0, \quad \textrm{for all } 0< t < T, 
\end{equation}    
with the initial and terminal conditions $X(0)=x$ and $X(T)=0$. The solution to (\ref{ode}) is (\ref{stat-min}). By Theorem \ref{thm-opt-stat} this is the unique minimizer of (\ref{v-cost}). 

\end{appendix}

  \end{document}